\documentclass[submission,copyright,creativecommons]{eptcs}
\providecommand{\event}{HSB 2013} 
\usepackage{color}
\usepackage{amsmath}
\usepackage{amsfonts}
\usepackage{amsthm}
\usepackage{graphicx}
\usepackage{pifont}
\usepackage{array, color}
\usepackage{booktabs}
\usepackage{amssymb}
\usepackage{mathrsfs}
\usepackage{subfigure}
\usepackage{setspace}
\usepackage{epstopdf}
\usepackage[normalem]{ulem}
\usepackage{pstool} 

\title{A Hybrid Model of a Genetic Regulatory Network\\ in Mammalian Sclera}
\author{Qin Shu 
\institute{Department of Aerospace and Mechanical Engineering\\
University of Arizona, Tucson, Arizona, USA}
\email{shuq@email.arizona.edu}
\and 
Diana Catalina Ardila
\institute{Department of Aerospace and Mechanical Engineering\\
University of Arizona, Tucson, Arizona, USA}
\email{cardila@email.arizona.edu}
\and
Ricardo G. Sanfelice
\institute{Department of Aerospace and Mechanical Engineering\\
University of Arizona, Tucson, Arizona, USA}
\email{sricardo@u.arizona.edu}
\and
Jonathan P. Vande Geest
\institute{Department of Aerospace and Mechanical Engineering\\
Biomedical Engineering, BIO5 Institute\\
University of Arizona, Tucson, Arizona, USA}
\email{jpv1@email.arizona.edu}
}

\def\titlerunning{A Hybrid Model of a Genetic Regulatory Network in Mammalian Sclera}
\def\authorrunning{Q. Shu, D. C. Ardila, R. G. Sanfelice, and J. P. Vande Geest}
\begin{document}
\maketitle

\vspace{-0.1in}
\begin{abstract}
Myopia in human and animals is caused by the axial elongation of the eye and is closely linked to the thinning of the sclera which supports the eye tissue. This thinning has been correlated with the overproduction of matrix metalloproteinase (MMP-2), an enzyme that degrades the collagen structure of the sclera. In this short paper, we propose a descriptive model of a regulatory network with hysteresis, which seems necessary for creating oscillatory behavior in the hybrid model between MMP-2, MT1-MMP and TIMP-2.  Numerical results provide insight on the type of equilibria present in the system.
\end{abstract}

\vspace{-0.1in}
\section{Introduction}

This short paper presents a descriptive model of a genetic regulatory network in the mammalian sclera using the formalism of hybrid dynamical systems. This model is deduced from experimental observations of enzyme interactions that govern the remodeling of the collagen tissue in the sclera. A number of research publications indicate that myopia is closely related to an unbalanced remodeling in sclera \cite{99,98}. Myopia is an optical condition in which the eye grows abnormally in the axial direction, causing images to form in front of the retina compared to on the retina, as it normally occurs \cite{ji2009form,99,mcbrien2000scleral,Rada2006sclera}. The excessive length of the eye is driven by the remodeling of the scleral extra cellular matrix (ECM) (e.g., loss of Type I collagen, COL1A1), leading the progressive thinning of this tissue \cite{backhouse2010effect,mcbrien2013regulation,99}. Scleral remodeling is regulated by a large number of growth factors, membrane receptors, proteases, and protease inhibitors, which work in concert to optimize the dynamic synthesis and degradation of COL1A1\cite{gentle2003collagen,mcbrien2013regulation,yang2009myopia}. One of the most studied actors in sclera remodeling is the Type II matrix metalloproteinase (MMP-2), because of its role in the degradation of COL1A1 \cite{93,99,94}.
MMP-2 is regulated by the Type II tissue inhibitor of the matrix metalloproteinases (TIMP-2), and when the two enzymes are properly balanced, the sclera develops normally. MMP-2 regulation by TIMP-2 shows a particular mechanism in which TIMP-2 not only inhibits the proteolytic activity of MMP-2, but is also necessary for the production of this metalloproteinase in its active form \cite{99,siegwart2005selective, xiong2006effects}. Such a mechanism is very important for the balance between COL1A1 production and degradation in sclera, and hence, should play a key role in a model of a genetic regulatory network in this tissue.
   
The remainder of the paper is organized as follows. Section~\ref{sec:modeling} introduces the mechanisms governing the regulatory network of interest and proposes a hybrid system model. Section~\ref{sec:simulations} presents results from simulations of the proposed model, which, for a particular set of parameters, identify both isolated equilibria and limit cycles. Final remarks and a discussion of the current efforts appear in Section~\ref{sec:conclusion}.

\section{Modeling}
\label{sec:modeling}

\begin{figure}[h!]
 \centering
  \psfragfig*[width=1\columnwidth]{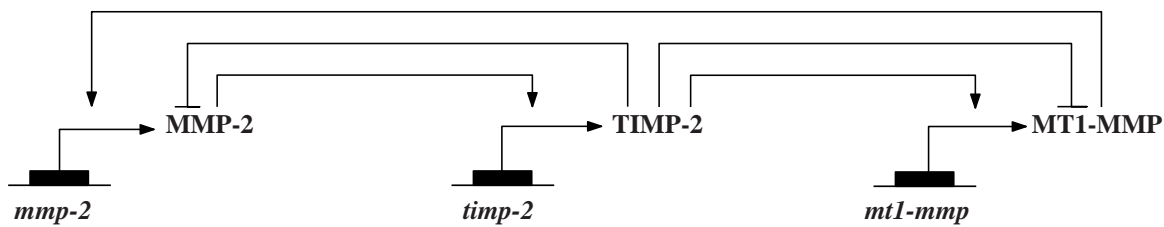}  
  {
  \psfrag{gen1}{\small\textbf{\textit{mmp-2}}}
  \psfrag{gen2}{\small\textbf{\textit{timp-2}}}
  \psfrag{gen3}{\small\textbf{\textit{mt1-mmp}}}
  \psfrag{prot1}{\small\textbf{MMP-2}}
  \psfrag{prot2}{\small\textbf{TIMP-2}}
  \psfrag{prot3}{\small\textbf{MT1-MMP}}  
  }
  \vspace{-0.4in}
  \caption{\emph{Proposed genetic regulatory network for sclera. Lowercase names refer to genes, uppercase names refer to proteins. Lines ending in arrows represent expression triggers and lines ending in flatheads refer to inhibition triggers.}}
\label{fig:1}
\end{figure}

We develop a model of a regulatory network in mammalian sclera from the following experimental observations.  
Sufficient high levels of MMP-2 protein cause the expression of  TIMP-2 \cite{99,95} (considering expression as the result of transcription, translation and activation of the protein latent form). When the concentration of TIMP-2 exceeds a minimum threshold, this protein indirectly modulates the increment of MMP-2: TIMP-2 triggers the expression of active membrane-type I matrix metalloproteinase (MT1-MMP) \cite{99,95,siegwart2005selective}, which is necessary for the activation of latent MMP-2 \cite{99,siegwart2005selective,xiong2006effects}. 
When the concentration of TIMP-2 protein is sufficiently high, TIMP-2 inhibits the proteolytic activity of MMP-2 and MT1-MMP \cite{99,sakalihasan1996activated,95,siegwart2005selective,xiong2006effects}. As we mentioned above, MT1-MMP triggers the activation of latent MMP-2
when sufficiently high \cite{guo2005paradigmatic,monea2002plasmin};
therefore, by blocking MT1-MMP, TIMP-2 is also inhibiting the activation of  latent MMP-2 \cite{99,siegwart2005selective,xiong2006effects}.  In fact, \cite{99,xiong2006effects} argue that the increased TIMP-2 mRNA and protein levels are significant as TIMP-2 is not only a protein inhibitor of both the active and latent form of MMP-2 but also paradoxically essential for the MT1-MMP dependent activation of MMP-2. 
The genetic network capturing these mechanisms is depicted in Figure~\ref{fig:1}.

The mechanisms described above can be encoded in a piecewise-linear differential equation following the modeling technique in \cite{93,82}.
However, the resulting model of the genetic network in sclera
would not incorporate hysteresis, which is a key player in genetic regulatory networks \cite{91,89,90,92}.
To incorporate hysteresis, 
we follow the approach in \cite{Shu.Sanfelice.13.I&C} and propose a hybrid system model in the framework of \cite{83}.
To this end, we define the state of the hybrid system as
\begin{equation}
z=[ x_1, x_2, x_3, q_1, q_2, q_3, q_4]^\top
\end{equation}
where $z\in Z:=\mathbb{R}^{3}_{\geq0}\times\{0, 1\}^4$.
The continuous states  $x_1, x_2, x_3$ represent the protein concentrations, 
where $x_1$ represents the protein concentration of TIMP-2, 
$x_2$ the concentration of MT1-MMP, and 
$x_3$ the concentration of MMP-2.
Positive constants $\gamma_1, \gamma_2, \gamma_3$ define the decay rates
and $k_1, k_2, k_3$ define the growth rates, respectively, for each of the concentrations.
The discrete states (logic variables)
$q_1, q_2, q_3, q_4$ define the boolean value ($1$ or $0$) of the hysteresis functions associated with each of the thresholds $\theta_1, \theta_2, \theta_3, \theta_4$
and the hysteresis half-width constants
$h_1, h_2, h_3, h_4$ associated with each of the thresholds, respectively.

\begin{table}[ht]
\begin{center}
\begin{tabular}{|c|c|}
\hline
Threshold & Definition  \\\hline
$\theta_1 $& TIMP-2 level for MT1-MMP expression \\\hline
$\theta_2 $& MT1-MMP level for MMP-2 expression\\\hline
$\theta_3$& TIMP-2 level for MT1-MMP/MMP-2 inhibition\\\hline
$\theta_4$ & MMP-2 level for TIMP-2 expression\\
\hline
\end{tabular}
\end{center}
\caption{\label{tab:D}Definition of protein thresholds in the genetic regulatory network for sclera. }
\vspace{-0.35in}
\end{table}

Following the definitions in Table~\ref{tab:D}, the thresholds and hysteresis half-width constants are used to determine when, 
for current values of the protein concentrations and of the logic variables,
changes of the logic variables should occur.
For instance, according to the mechanisms described above,
if $q_4 = 0$ and 
$x_3$ is small, then $x_1$ should decay according to its decay rate $\gamma_1$.  
However, if $q_4 = 0$ and $x_3$ becomes large (i.e., the concentration of MMP-2 is large) then 
$q_4$ should change to $1$ and $x_1$ should be expressed according to its own growth rate $k_1$.
The continuous evolution of $x_3$ can be captured mathematically by the differential equation
$$
\dot{x}_1 = k_1 q_4 - \gamma_1 x_1
$$
while the discrete change of $q_4$ can be captured by the difference equation
$$
q_4^+ = 1-q_4 \qquad 
\mbox{ when }\qquad
q_4 = 0 \mbox{ and } x_3 \geq \theta_4+h_4,\qquad \mbox{or} \quad
q_4 = 1 \mbox{ and } x_3 \leq \theta_4-h_4
$$
In this way, the {\em flow map} of the hybrid system defining the continuous dynamics of $z$ is given by
\begin{equation}
F(z):=\left[
\begin{array}{ll}
k_1q_4-\gamma_1x_1\\
k_2q_1(1-q_3)-\gamma_2x_2\\
k_3q_2(1-q_3)-\gamma_3x_3\\
0_{4\times 1}\end{array}
\right]
\end{equation}
Changes of the variables occur when $z$ is in the {\em jump set}, which is conveniently written as
$$D:=\bigcup_{i=1}^4 D_i$$
where
$$D_1:=\{z: q_1=1, x_1\leq\theta_1-h_1\}\cup\{z: q_1=0, x_1\geq\theta_1+h_1\}$$
$$D_2:=\{z: q_2=1, x_2\leq\theta_2-h_2\}\cup\{z: q_2=0, x_2\geq\theta_2+h_2\}$$
$$D_3:=\{z: q_3=1, x_1\leq\theta_3-h_3\}\cup\{z: q_3=0, x_1\geq\theta_3+h_3\}$$
$$D_4:=\{z: q_4=1, x_3\leq\theta_4-h_4\}\cup\{z: q_4=0, x_3\geq\theta_4+h_4\}$$
\indent\setlength\parindent{2em}The right-hand side of the difference equation discretely updating the logic variables
is given by the {\em jump map}
\begin{equation}G(z)=
\begin{cases}
g_1(z)&z\in D_1\setminus(D_2\cup D_3\cup D_4)\\
g_2(z)&z\in D_2\setminus(D_1\cup D_3\cup D_4)\\
g_3(z)&z\in D_3\setminus(D_1\cup D_2\cup D_4)\\
g_4(z)&z\in D_4\setminus(D_1\cup D_2\cup D_3)\\
\hat{g}(z)&z\in D_1\cap D_2\cap D_3\cap D_4
\end{cases}
\end{equation}
where 
$$g_1(z)=\left[
\begin{array}{l}
x_1\\x_2\\x_3\\1-q_1\\q_2\\q_3\\q_4
\end{array}\right],\quad g_2(z)=\left[
\begin{array}{l}
x_1\\x_2\\x_3\\q_1\\1-q_2\\q_3\\q_4
\end{array}\right],\quad g_3(z)=\left[
\begin{array}{l}
x_1\\x_2\\x_3\\q_1\\q_2\\1-q_3\\q_4
\end{array}\right],\quad
g_4(z)=\left[
\begin{array}{l}
x_1\\x_2\\x_3\\q_1\\q_2\\q_3\\1-q_4
\end{array}\right]$$
and
$\hat{g}(z)=\{g_1(z), g_2(z), g_3(z), g_4(z)\}$.
Note that $x_1$ and its associated logic variables $q_1$ and $q_3$ 
are the only ``inputs'' to the dynamics of $x_2$, which suggests that
$\theta_1+h_1<\theta_3-h_3$ should hold for $x_2$ to ever grow.
Moreover, $x_1$ and its associated logic variable $q_3$ are ``inputs''
to the dynamics of $x_3$, while $x_3$ and $q_4$ are ``inputs'' 
to the dynamics of $x_1$, in what resembles to a feedback interconnection.

With the definitions mentioned before, a hybrid system ${\cal H}=(F, C, G, D)$ 
in the framework of \cite{83}
capturing the mechanism in the genetic network of sclera with hysteresis is given as
\begin{equation}
{\cal H}:\quad z\in Z\quad \left\{
\begin{array}{ll}
\ \ \dot{z}=F(z)&\qquad\textrm{$z\in C:=\overline{Z\setminus D}$}\\
z^+\in G(z)&\qquad\textrm{$z\in D$}
\end{array}\right.
\label{eqn:H}
\end{equation}

\section{Simulation Results}
\label{sec:simulations}

We simulate the hybrid model of the scleral genetic network within a Matlab/Simulink toolbox \cite{92}. 
Unless otherwise stated, the growth rate $k_i$ and decay rates $\gamma_i$, $i=1,2,3$ for the three proteins are identically set to $1$ and the hysteresis $h_i$, $i=1, 2, 3, 4$ are set to $0.01$.

\subsection{Isolated Equilibrium Points}

\begin{figure}[h!]
\begin{center}
\subfigure[$\theta_1=0.4,\theta_2=0.5, \theta_3=0.6, \theta_4=0.7, x_1(0)=0.15, x_2(0)=0.45, x_3(0)=0.8, q_1(0)=1, q_2(0)=1, q_3(0)=0, q_4(0)=1,$ $*$ is the initial point.\label{fig:s1}]
{\includegraphics[width=0.45\columnwidth]{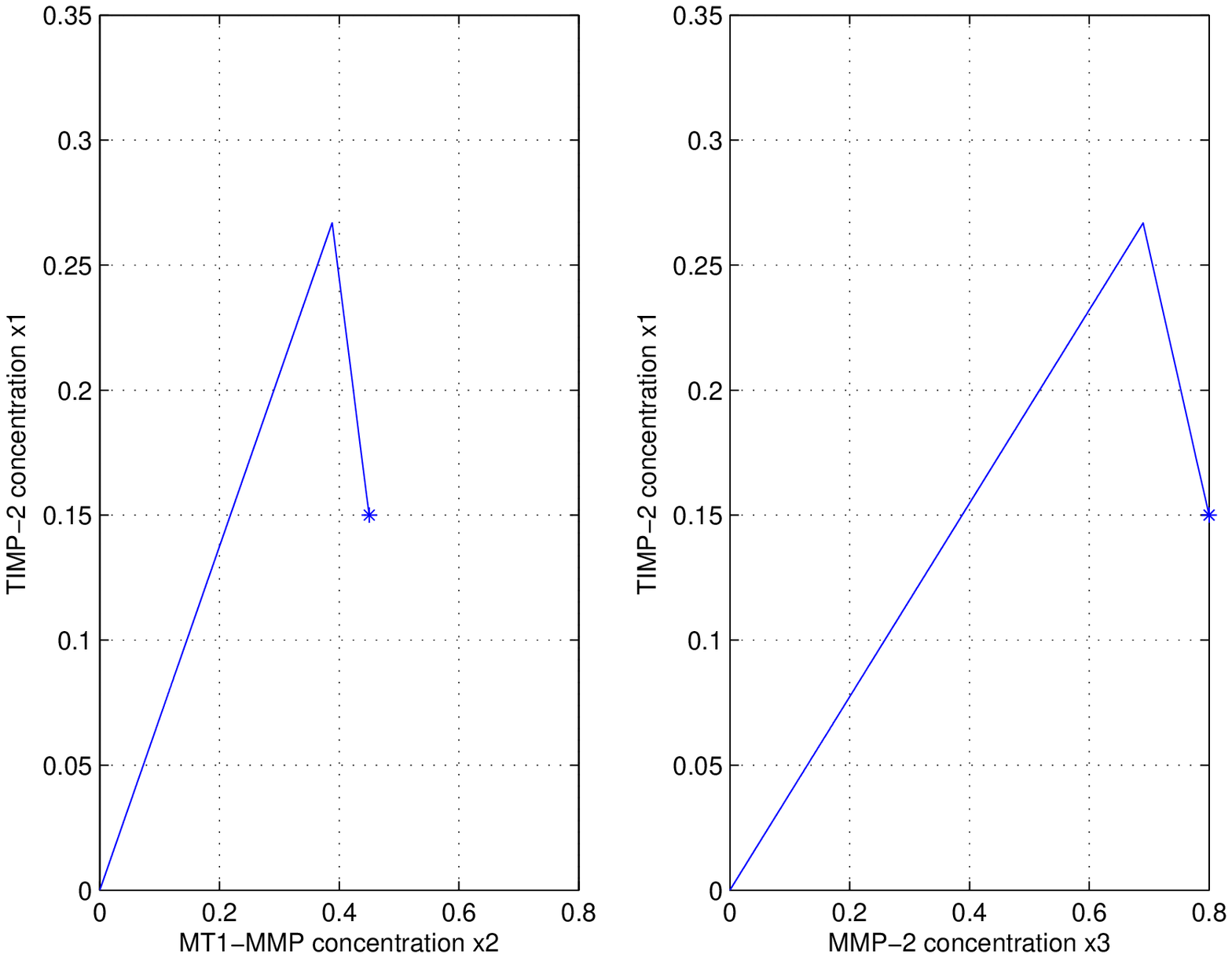}}\qquad
\subfigure[$\theta_1=0.4, \theta_2=0.5, \theta_3=0.6, \theta_4=0.7, x_1(0)=0.15, x_2(0)=0.45, x_3(0)=0.8, q_1(0)=1, q_2(0)=1, q_3(0)=0, q_4(0)=1,$ $*$ is the initial point.\label{fig:s2}]
{\includegraphics[width=0.45\columnwidth]{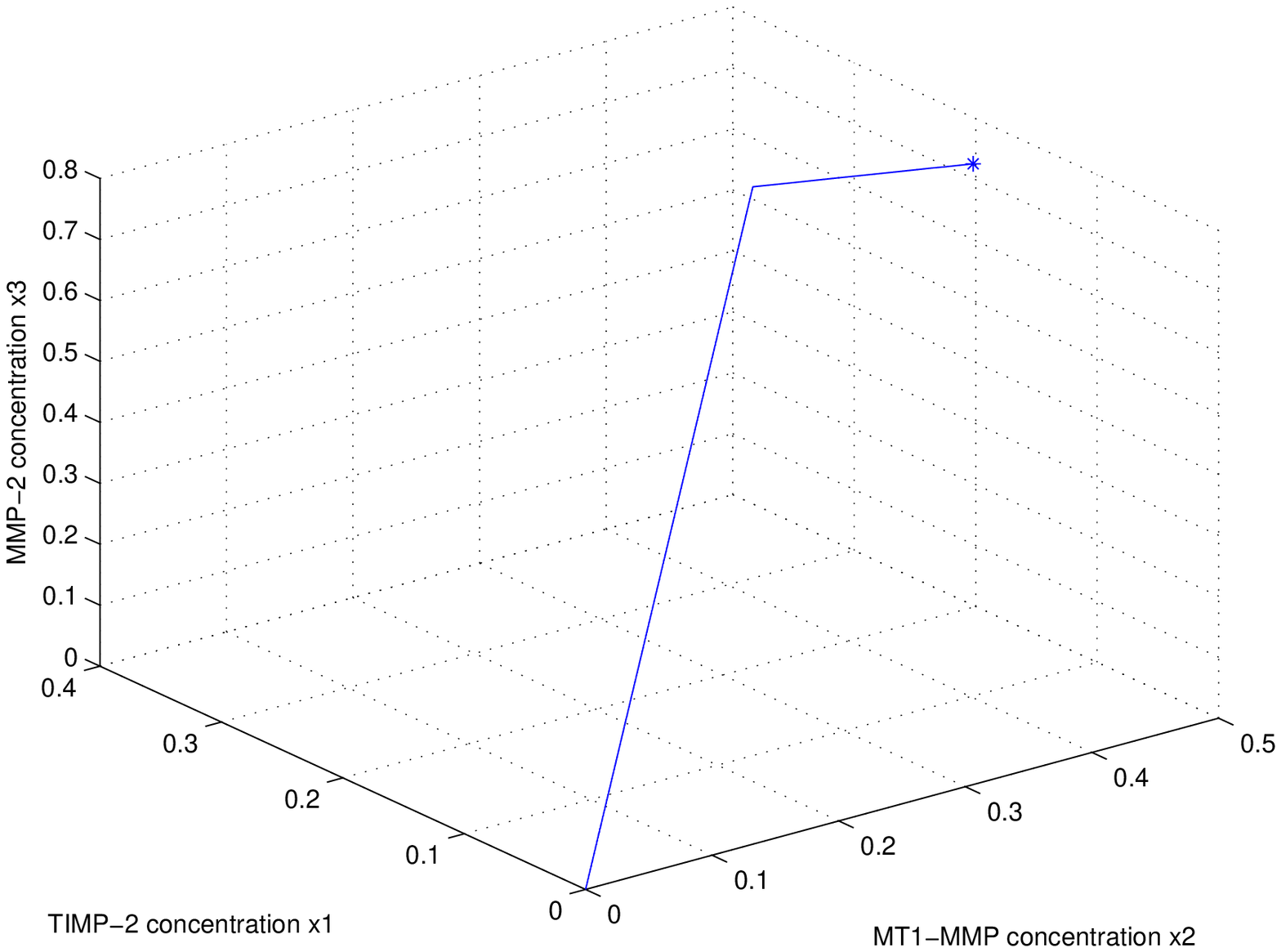}}\qquad
\subfigure[$k_1=0.55, k_3=0.9, \theta_1=0.4, \theta_2=0.5, \theta_3=0.6, \theta_4=0.7, x_1(0)=0.45, x_2(0)=0.6, x_3(0)=0.8, q_1(0)=1, q_2(0)=1, q_3(0)=0, q_4(0)=1,$ $*$ is the initial point.\label{fig:s3}]
{\includegraphics[width=0.45\columnwidth]{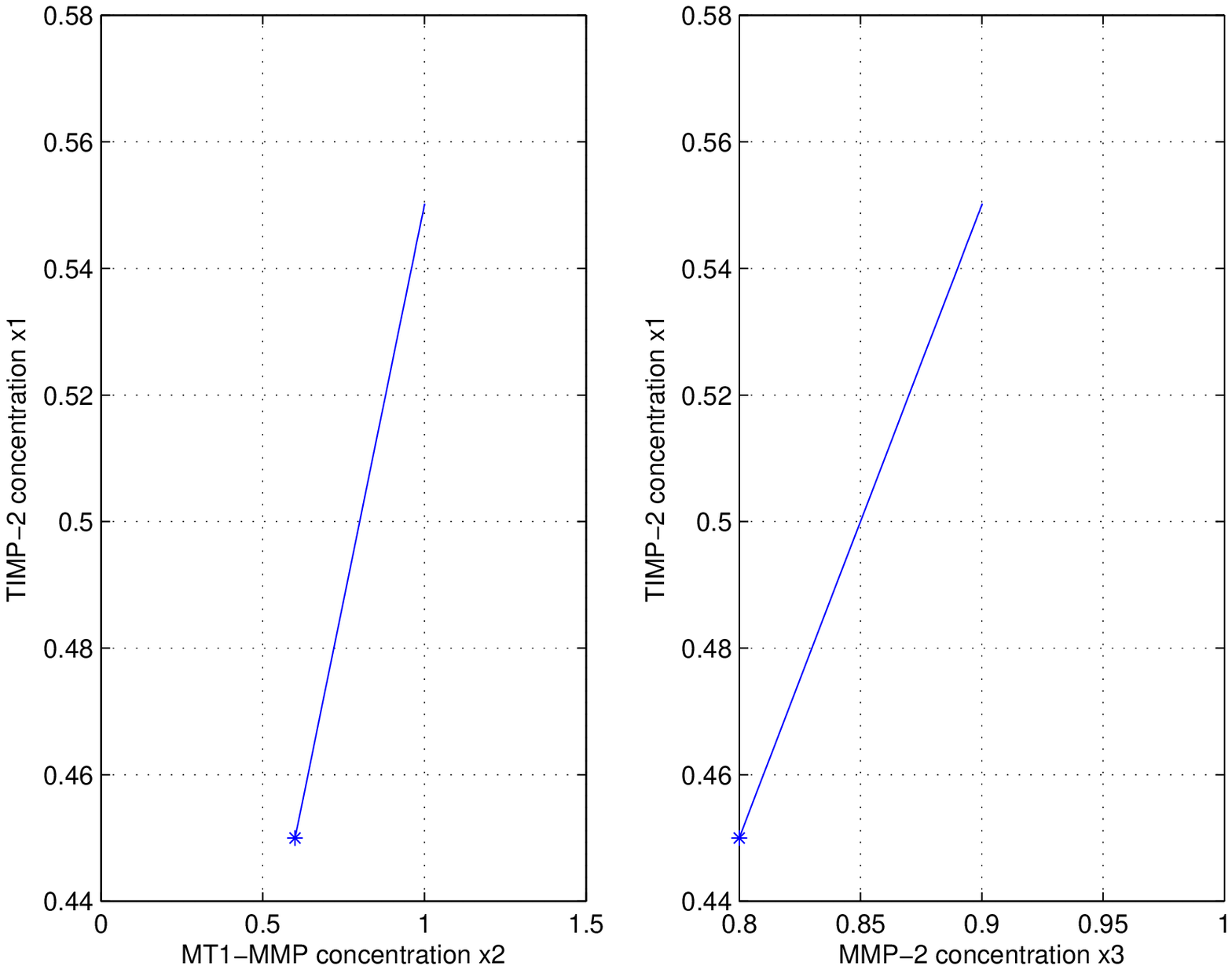}}\qquad
\subfigure[$k_1=0.55, k_3=0.9, \theta_1=0.4, \theta_2=0.5, \theta_3=0.6, \theta_4=0.7, x_1(0)=0.45, x_2(0)=0.6, x_3(0)=0.8, q_1(0)=1, q_2(0)=1, q_3(0)=0, q_4(0)=1,$ $*$ is the initial point.\label{fig:s4}]
{\includegraphics[width=0.45\columnwidth]{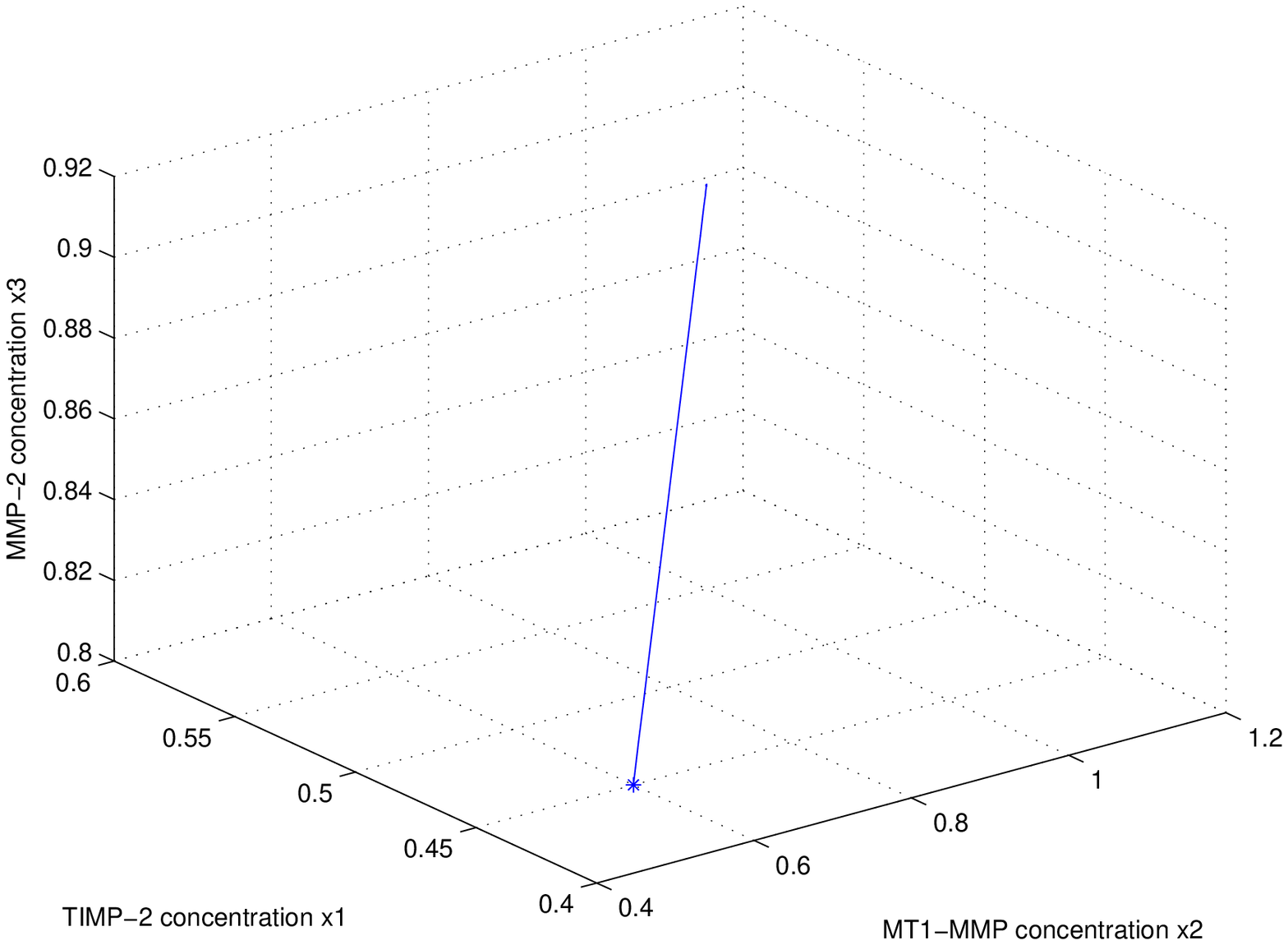}}
\end{center}
\caption{Solutions to $\cal H$ for different parameters and initial conditions. For the chosen values, solutions converge to isolated equilibrium points.}
\end{figure}

Figure~\ref{fig:s1} and Figure~\ref{fig:s2} present simulation results in which the hybrid system evolves to the equilibrium point at $x^*=(0, 0, 0)$. Under these initial conditions and protein thresholds, the concentration of TIMP-2 ($x_1$) is not sufficiently high to permit continued expression of the MT1-MMP ($x_2$) and MMP-2 ($x_3$) genes. The protein concentration associated with the MMP-2 gene continues to grow, but when the MT1-MMP gene is inhibited, MMP-2 will become inhibited with time.

Figure~\ref{fig:s3} and Figure~\ref{fig:s4} show that the solution of the hybrid system goes toward the equilibrium point at $x^*=\left(\frac{k_1}{\gamma_1}, \frac{k_2}{\gamma_2}, \frac{k_3}{\gamma_3}\right)$. With the given initial conditions and parameters, the concentration of TIMP-2 ($x_1$) is not high enough to inhibit the expression of the MT1-MMP ($x_2$) and MMP-2 ($x_3$) genes. This situation can be a cause of high myopia \cite{99, rada1999gelatinase}. 

\subsection{Limit cycles}

\begin{figure}[h!]
\begin{center}
\subfigure[ $h_i = 0.01$ for each $i=1,2,3,4$, $*$ is the initial point. \label{fig:s5}]
{\includegraphics[width=0.45\columnwidth]{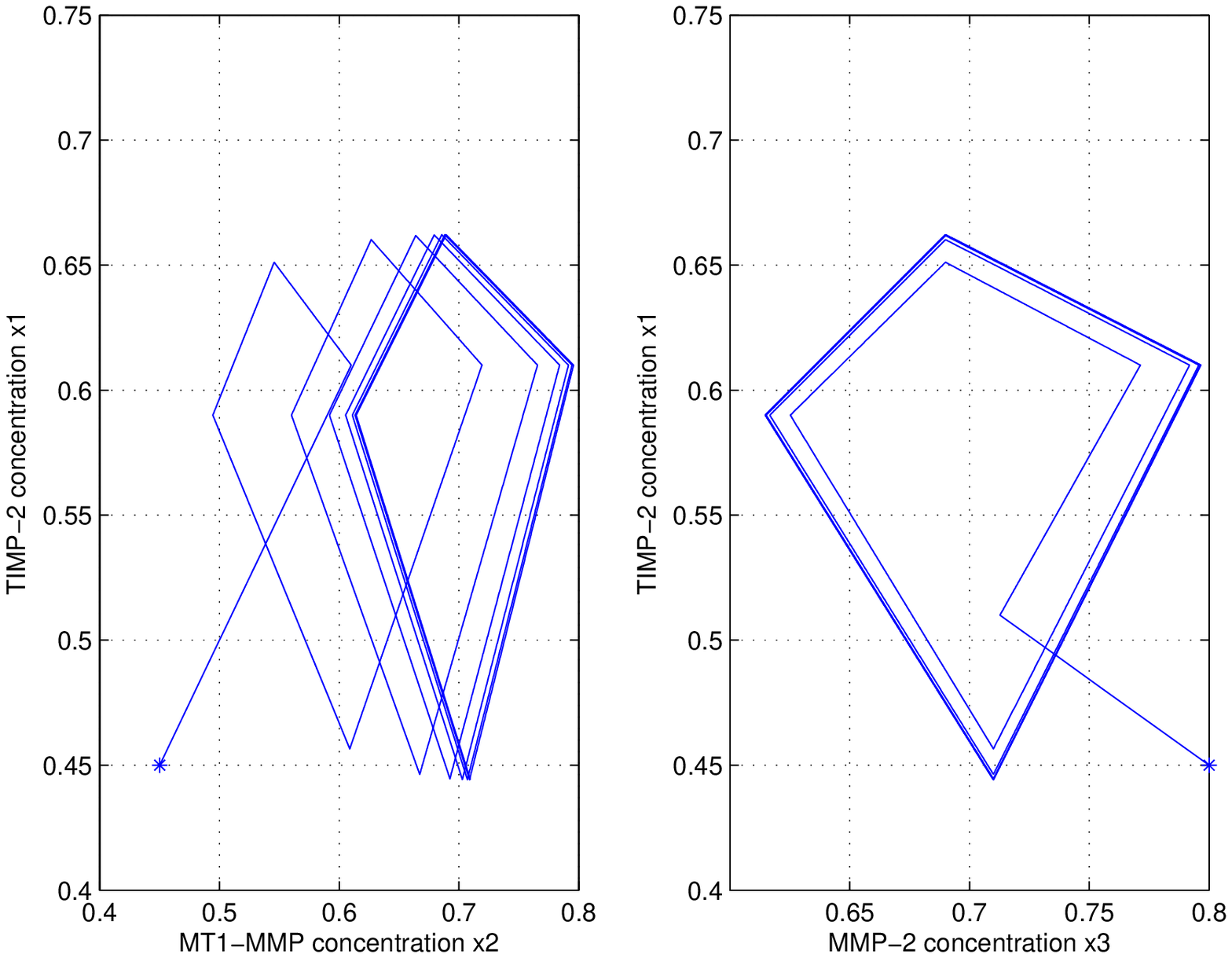}}\qquad
\subfigure[$h_i = 0.01$ for each $i=1,2,3,4$, $*$ is the initial point. \label{fig:s6}]
{\includegraphics[width=0.5\columnwidth]{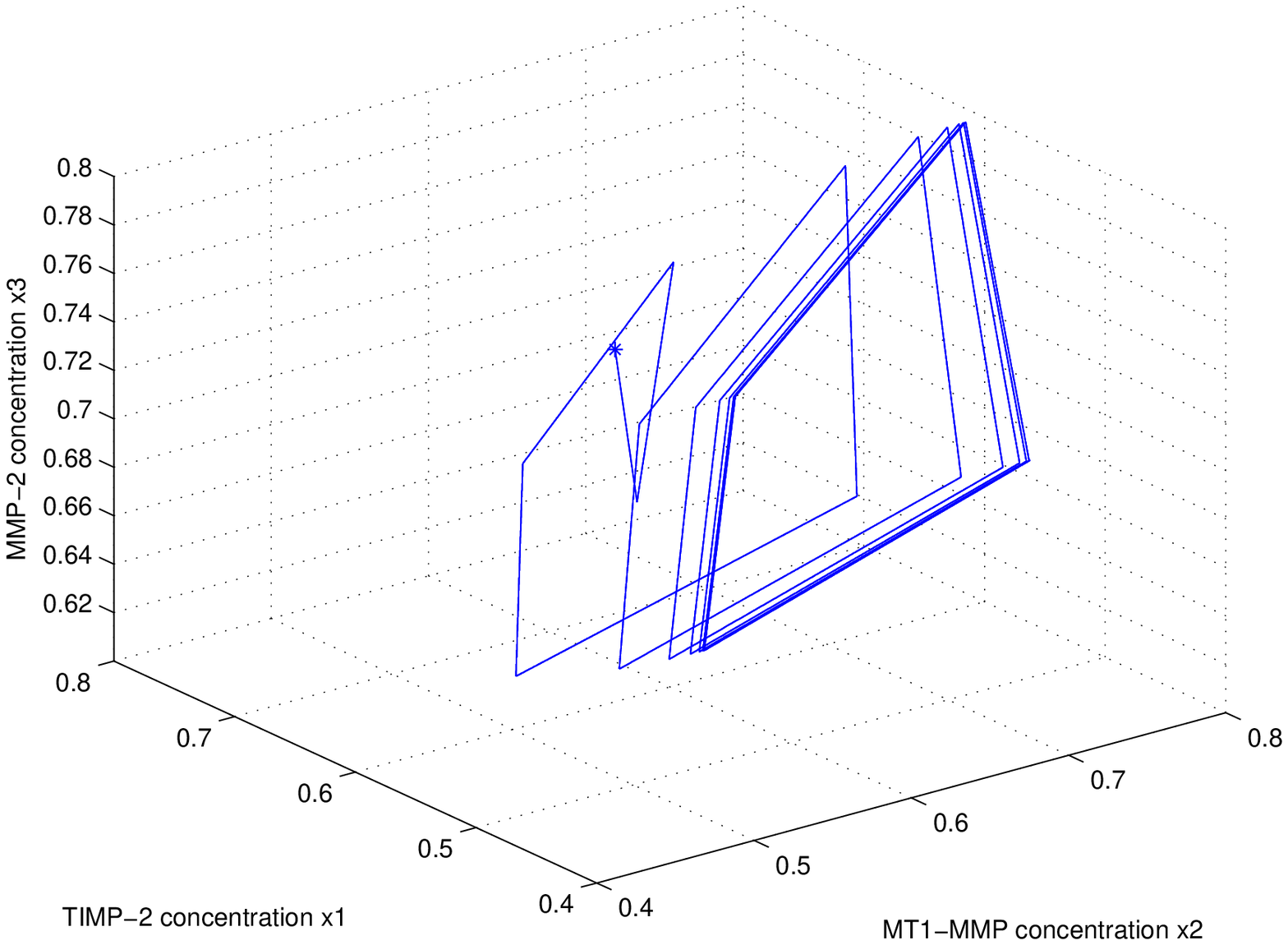}}
\subfigure[$h_i = 0$ for each $i=1,2,3,4$, $*$ is the initial point. \label{fig:s7}]
{\includegraphics[width=0.45\columnwidth]{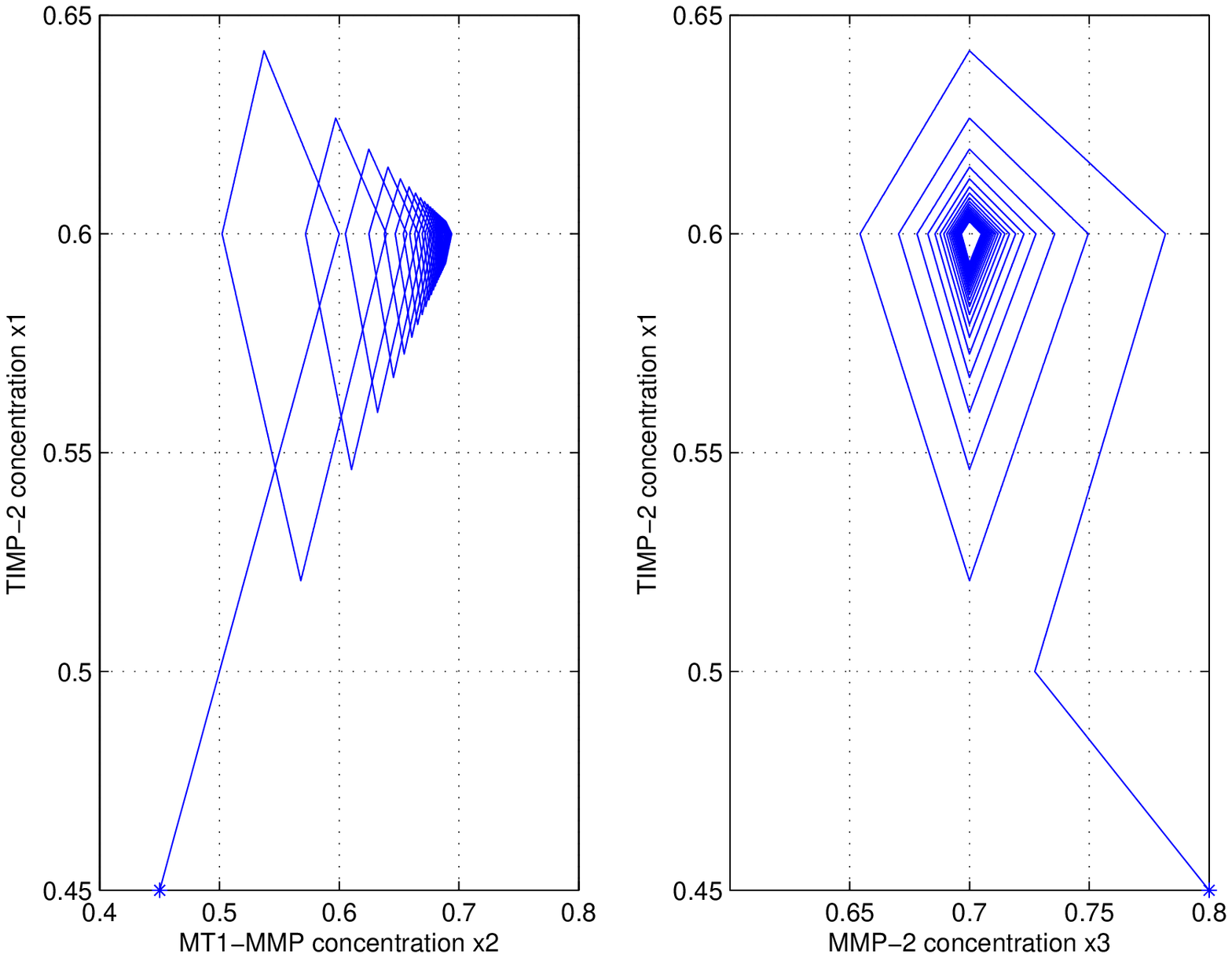}}\qquad
\subfigure[$h_i = 0$ for each $i=1,2,3,4$, $*$ is the initial point. \label{fig:s8}]
{\includegraphics[width=0.45\columnwidth]{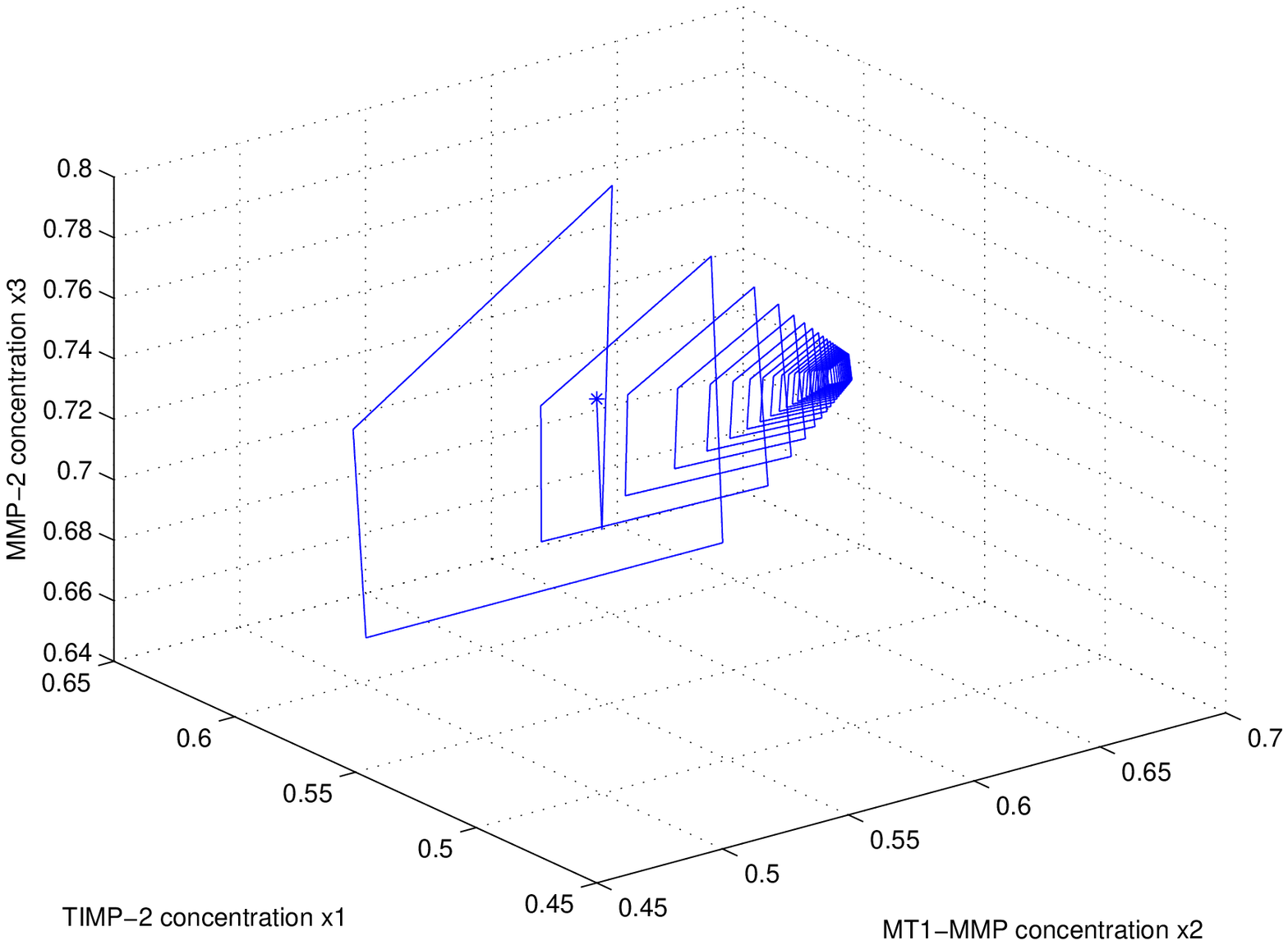}}
\end{center}
\vspace{-0.2in}
\caption{Solutions to $\cal H$ for $\theta_1=0.4, \theta_2=0.5, \theta_3=0.6, \theta_4=0.7, x_1(0)=0.45, x_2(0)=0.45, x_3(0)=0.8, q_1(0)=1, q_2(0)=1, q_3(0)=0, q_4(0)=1$ with nonzero and zero $h_i$ for each $i=1,2,3,4$.}
\end{figure}

Figure~\ref{fig:s5} and Figure~\ref{fig:s6} illustrates the oscillatory behavior in the hybrid system when the concentration of TIMP-2 exceeds $\theta_1$ and the concentration of MMP-2 exceeds $\theta_4$ recurrently. In this scenario, the discrete state behavior stabilizes to a periodic orbit. It is apparent that the TIMP-2 protein as modeled here has a stabilizing effect on the other two protein concentrations when it is at a sufficiently high level. In this scenario, the sclera develops normally. To illustrate that such normal development of the sclera is only possible when hysteresis is present, the previous simulation is repeated for half-width hysteresis constants equal to zero. Figure~ \ref{fig:s7} and Figure~\ref{fig:s8} show the corresponding system response. The solution to the hybrid system converges to an isolated equilibrium point.

\section{Conclusion}
\label{sec:conclusion}

A mathematical model of a regulatory network with hysteresis to describe the mechanisms
in the mammalian sclera was introduced.  The model captures the interaction between MMP-2, MT1-MMP, and TIMP-2. Numerical results indicate that the system can have both isolated equilibria and limit cycles in the 3-dimensional space of protein concentrations. 
For the arbitrarily chosen parameters, numerical results seem to suggest that hysteresis is needed for normal 
development of sclera.
Current efforts include characterizing the type of equilibria in terms of the 
values of the systems constants using the hybrid systems techniques employed in
\cite{Shu.Sanfelice.13.I&C} and the design of in-vivo experiments to identify the parameters of the
genetic model.

\vspace{-0.1in}
\section{Acknowledgments}

Research by D. C. Ardila and J. P. Vande Geest has been partially supported by the NIH research grant 1RO11EY020890-02A1. Research by R. G. Sanfelice has been partially supported by the National Science Foundation under CAREER Grant no. ECS-1150306 and by the Air Force Office of Scientific Research under YIP Grant no. FA9550-12-1-0366.

\vspace{-0.2in}
\nocite{*}
\bibliographystyle{eptcs}
\bibliography{myopia}

\begin{thebibliography}{10}
\providecommand{\bibitemdeclare}[2]{}
\providecommand{\surnamestart}{}
\providecommand{\surnameend}{}
\providecommand{\urlprefix}{Available at }
\providecommand{\url}[1]{\texttt{#1}}
\providecommand{\href}[2]{\texttt{#2}}
\providecommand{\urlalt}[2]{\href{#1}{#2}}
\providecommand{\doi}[1]{doi:\urlalt{http://dx.doi.org/#1}{#1}}
\providecommand{\bibinfo}[2]{#2}

\bibitemdeclare{article}{backhouse2010effect}
\bibitem{backhouse2010effect}
\bibinfo{author}{S.~\surnamestart Backhouse\surnameend} \&
  \bibinfo{author}{J.~R. \surnamestart Phillips\surnameend}
  (\bibinfo{year}{2010}): \emph{\bibinfo{title}{Effect of induced myopia on
  scleral myofibroblasts and in vivo ocular biomechanical compliance in the
  Guinea pig}}.
\newblock {\sl \bibinfo{journal}{Investigative Ophthalmology \& Visual
  Science}} \bibinfo{volume}{51}(\bibinfo{number}{12}), pp.
  \bibinfo{pages}{6162--6171}, \doi{10.1167/iovs.10-5387}.

\bibitemdeclare{article}{91}
\bibitem{91}
\bibinfo{author}{J.~\surnamestart Das\surnameend},
  \bibinfo{author}{M.~\surnamestart Ho\surnameend},
  \bibinfo{author}{J.~\surnamestart Zikherman\surnameend},
  \bibinfo{author}{C.~\surnamestart Govern\surnameend},
  \bibinfo{author}{M.~\surnamestart Yang\surnameend},
  \bibinfo{author}{A.~\surnamestart Weiss\surnameend}, \bibinfo{author}{A.~K.
  \surnamestart Chakraborty\surnameend} \& \bibinfo{author}{J.~P. \surnamestart
  Roose\surnameend} (\bibinfo{year}{2009}): \emph{\bibinfo{title}{Digital
  Signaling and Hysteresis Characterize Ras Activation in Lymphoid Cells}}.
\newblock {\sl \bibinfo{journal}{Cells}} \bibinfo{volume}{136}, pp.
  \bibinfo{pages}{337--351}, \doi{10.1016/j.cell.2008.11.051}.

\bibitemdeclare{article}{gentle2003collagen}
\bibitem{gentle2003collagen}
\bibinfo{author}{A.~\surnamestart Gentle\surnameend},
  \bibinfo{author}{Y.~\surnamestart Liu\surnameend}, \bibinfo{author}{J.~E.
  \surnamestart Martin\surnameend}, \bibinfo{author}{G.~L. \surnamestart
  Conti\surnameend} \& \bibinfo{author}{N.~A. \surnamestart McBrien\surnameend}
  (\bibinfo{year}{2003}): \emph{\bibinfo{title}{Collagen gene expression and
  the altered accumulation of scleral collagen during the development of high
  myopia}}.
\newblock {\sl \bibinfo{journal}{Journal of Biological Chemistry}}
  \bibinfo{volume}{278}(\bibinfo{number}{19}), pp.
  \bibinfo{pages}{16587--16594}, \doi{10.1074/jbc.M300970200}.

\bibitemdeclare{book}{83}
\bibitem{83}
\bibinfo{author}{R.~\surnamestart Goebel\surnameend}, \bibinfo{author}{R.~G.
  \surnamestart Sanfelice\surnameend} \& \bibinfo{author}{A.~R. \surnamestart
  Teel\surnameend} (\bibinfo{year}{2012}): \emph{\bibinfo{title}{Hybrid
  Dynamical Systems: Modeling, Stability, and Robustness}}.
\newblock \bibinfo{publisher}{Princeton University Press}.

\bibitemdeclare{article}{guo2005paradigmatic}
\bibitem{guo2005paradigmatic}
\bibinfo{author}{C.~\surnamestart Guo\surnameend},
  \bibinfo{author}{J.~\surnamestart Jiang\surnameend},
  \bibinfo{author}{J.~\surnamestart Martin~Elliott\surnameend} \&
  \bibinfo{author}{L.~\surnamestart Piacentini\surnameend}
  (\bibinfo{year}{2005}): \emph{\bibinfo{title}{Paradigmatic identification of
  MMP-2 and MT1-MMP activation systems in cardiac fibroblasts cultured as a
  monolayer}}.
\newblock {\sl \bibinfo{journal}{Journal of cellular biochemistry}}
  \bibinfo{volume}{94}(\bibinfo{number}{3}), pp. \bibinfo{pages}{446--459},
  \doi{10.1002/jcb.20272}.

\bibitemdeclare{article}{93}
\bibitem{93}
\bibinfo{author}{J.~\surnamestart Hu\surnameend},
  \bibinfo{author}{D.~\surnamestart Cui\surnameend},
  \bibinfo{author}{X.~\surnamestart Yang\surnameend},
  \bibinfo{author}{S.~\surnamestart Wang\surnameend},
  \bibinfo{author}{S.~\surnamestart Hu\surnameend},
  \bibinfo{author}{C.~\surnamestart Li\surnameend} \&
  \bibinfo{author}{J.~\surnamestart Zeng\surnameend} (\bibinfo{year}{2008}):
  \emph{\bibinfo{title}{Bone morphogenetic protein-2: a potential regulator in
  scleral remodeling}}.
\newblock {\sl \bibinfo{journal}{Molecular Vision}} \bibinfo{volume}{14}, pp.
  \bibinfo{pages}{2370--2380}.

\bibitemdeclare{inproceedings}{89}
\bibitem{89}
\bibinfo{author}{J.~\surnamestart Hu\surnameend}, \bibinfo{author}{K.~R.
  \surnamestart Qin\surnameend}, \bibinfo{author}{C.~\surnamestart
  Xiang\surnameend} \& \bibinfo{author}{T.~H. \surnamestart Lee\surnameend}
  (\bibinfo{year}{2010}): \emph{\bibinfo{title}{Modeling of Hysteresis in a
  Mammalian Gene Regulatory Network}}.
\newblock In: {\sl \bibinfo{booktitle}{9th Annual International Conference on
  Computational Systems Bioinformatics}}, \bibinfo{volume}{9},
  \bibinfo{organization}{Life Sciences Society}, pp. \bibinfo{pages}{50--55}.

\bibitemdeclare{article}{96}
\bibitem{96}
\bibinfo{author}{M.~D. \surnamestart Jacobs\surnameend} (\bibinfo{year}{2009}):
  \emph{\bibinfo{title}{Multiscale systems integration in the eye}}.
\newblock {\sl \bibinfo{journal}{WIREs Systems Biology and Medicine}}
  \bibinfo{volume}{1}, pp. \bibinfo{pages}{15--27}, \doi{10.1002/wsbm.29}.

\bibitemdeclare{article}{ji2009form}
\bibitem{ji2009form}
\bibinfo{author}{F.~T. \surnamestart Ji\surnameend},
  \bibinfo{author}{Q.~\surnamestart Li\surnameend}, \bibinfo{author}{Y.~L.
  \surnamestart Zhu\surnameend}, \bibinfo{author}{L.~Q. \surnamestart
  Jiang\surnameend}, \bibinfo{author}{X.~T. \surnamestart Zhou\surnameend},
  \bibinfo{author}{M.~Z. \surnamestart Pan\surnameend} \&
  \bibinfo{author}{J.~\surnamestart Qu\surnameend} (\bibinfo{year}{2009}):
  \emph{\bibinfo{title}{Form Deprivation Myopia in C57BL/6 Mice}}.
\newblock {\sl \bibinfo{journal}{[Zhonghua yan ke za zhi] Chinese journal of
  ophthalmology}} \bibinfo{volume}{45}(\bibinfo{number}{11}), p.
  \bibinfo{pages}{1020}.

\bibitemdeclare{article}{90}
\bibitem{90}
\bibinfo{author}{B.~P. \surnamestart Kramer\surnameend} \&
  \bibinfo{author}{M.~\surnamestart Fussenegger\surnameend}
  (\bibinfo{year}{2005}): \emph{\bibinfo{title}{Hysteresis in a synthetic
  mammalian gene network}}.
\newblock {\sl \bibinfo{journal}{Proceedings of the National Academy of
  Sciences (USA)}} \bibinfo{volume}{102}, pp. \bibinfo{pages}{9517--9522},
  \doi{10.1073/pnas.0500345102}.

\bibitemdeclare{article}{mcbrien2013regulation}
\bibitem{mcbrien2013regulation}
\bibinfo{author}{N.~A. \surnamestart McBrien\surnameend}
  (\bibinfo{year}{2013}): \emph{\bibinfo{title}{Regulation of Scleral
  Metabolism in Myopia and the Role of Transforming Growth Factor-beta}}.
\newblock {\sl \bibinfo{journal}{Experimental eye research}},
  \doi{10.1016/j.exer.2013.01.014}.

\bibitemdeclare{article}{99}
\bibitem{99}
\bibinfo{author}{N.~A. \surnamestart McBrien\surnameend} \&
  \bibinfo{author}{A.~\surnamestart Gentile\surnameend} (\bibinfo{year}{2003}):
  \emph{\bibinfo{title}{Role of the sclera in the development and pathological
  complications of myopia}}.
\newblock {\sl \bibinfo{journal}{Progress in Retinal and Eye Research}}
  \bibinfo{volume}{22}, pp. \bibinfo{pages}{307--338},
  \doi{10.1016/S1350-9462(02)00063-0}.

\bibitemdeclare{article}{mcbrien2000scleral}
\bibitem{mcbrien2000scleral}
\bibinfo{author}{N.~A. \surnamestart McBrien\surnameend},
  \bibinfo{author}{P.~\surnamestart Lawlor\surnameend} \&
  \bibinfo{author}{A.~\surnamestart Gentle\surnameend} (\bibinfo{year}{2000}):
  \emph{\bibinfo{title}{Scleral Remodeling During the Development of and
  Recovery from Axial Myopia in the Tree Shrew}}.
\newblock {\sl \bibinfo{journal}{Investigative ophthalmology \& visual
  science}} \bibinfo{volume}{41}(\bibinfo{number}{12}), pp.
  \bibinfo{pages}{3713--3719}.

\bibitemdeclare{article}{82}
\bibitem{82}
\bibinfo{author}{T.~\surnamestart Mestl\surnameend},
  \bibinfo{author}{E.~\surnamestart Plahte\surnameend} \&
  \bibinfo{author}{S.~W. \surnamestart Omholt\surnameend}
  (\bibinfo{year}{1995}): \emph{\bibinfo{title}{A mathematical framework for
  describing and analysing gene regulatory networks}}.
\newblock {\sl \bibinfo{journal}{Journal of Theoretical Biology}}
  \bibinfo{volume}{176}, pp. \bibinfo{pages}{291--300},
  \doi{10.1006/jtbi.1995.0199}.

\bibitemdeclare{article}{monea2002plasmin}
\bibitem{monea2002plasmin}
\bibinfo{author}{S.~\surnamestart Monea\surnameend},
  \bibinfo{author}{K.~\surnamestart Lehti\surnameend},
  \bibinfo{author}{J.~\surnamestart Keski-Oja\surnameend} \&
  \bibinfo{author}{P.~\surnamestart Mignatti\surnameend}
  (\bibinfo{year}{2002}): \emph{\bibinfo{title}{Plasmin activates pro-matrix
  metalloproteinase-2 with a membrane-type 1 matrix metalloproteinase-dependent
  mechanism}}.
\newblock {\sl \bibinfo{journal}{Journal of cellular physiology}}
  \bibinfo{volume}{192}(\bibinfo{number}{2}), pp. \bibinfo{pages}{160--170},
  \doi{10.1002/jcp.10126}.

\bibitemdeclare{article}{94}
\bibitem{94}
\bibinfo{author}{E.~\surnamestart Morgunova\surnameend},
  \bibinfo{author}{A.~\surnamestart Tuuttila\surnameend},
  \bibinfo{author}{U.~\surnamestart Bergmann\surnameend} \&
  \bibinfo{author}{K.~\surnamestart Tryggvason\surnameend}
  (\bibinfo{year}{2002}): \emph{\bibinfo{title}{Structural insight into the
  complex formation of latent matrix metalloproteinase 2 with tissue inhibitor
  of metalloproteinase 2}}.
\newblock {\sl \bibinfo{journal}{Proceedings of the National Academy of
  Sciences of the USA}} \bibinfo{volume}{99}(\bibinfo{number}{11}), pp.
  \bibinfo{pages}{7414--7419}, \doi{10.1073/pnas.102185399}.

\bibitemdeclare{article}{rada1999gelatinase}
\bibitem{rada1999gelatinase}
\bibinfo{author}{J.~A. \surnamestart Rada\surnameend}, \bibinfo{author}{C.~A.
  \surnamestart Perry\surnameend}, \bibinfo{author}{M.~L. \surnamestart
  Slover\surnameend} \& \bibinfo{author}{V.~R. \surnamestart Achen\surnameend}
  (\bibinfo{year}{1999}): \emph{\bibinfo{title}{Gelatinase A and TIMP-2
  expression in the fibrous sclera of myopic and recovering chick eyes}}.
\newblock {\sl \bibinfo{journal}{Investigative ophthalmology \& visual
  science}} \bibinfo{volume}{40}(\bibinfo{number}{13}), pp.
  \bibinfo{pages}{3091--3099}.

\bibitemdeclare{article}{Rada2006sclera}
\bibitem{Rada2006sclera}
\bibinfo{author}{J.~A. \surnamestart Rada\surnameend},
  \bibinfo{author}{S.~\surnamestart Shelton\surnameend} \&
  \bibinfo{author}{T.~T. \surnamestart Norton\surnameend}
  (\bibinfo{year}{2006}): \emph{\bibinfo{title}{The Sclera and Myopia}}.
\newblock {\sl \bibinfo{journal}{Experimental eye research}}
  \bibinfo{volume}{82}(\bibinfo{number}{2}), pp. \bibinfo{pages}{185--200},
  \doi{10.1016/j.exer.2005.08.009}.

\bibitemdeclare{article}{sakalihasan1996activated}
\bibitem{sakalihasan1996activated}
\bibinfo{author}{N.~\surnamestart Sakalihasan\surnameend},
  \bibinfo{author}{P.~\surnamestart Delvenne\surnameend},
  \bibinfo{author}{B.~V. \surnamestart Nusgens\surnameend},
  \bibinfo{author}{R.~\surnamestart Limet\surnameend} \& \bibinfo{author}{C.~M.
  \surnamestart Lapi{\`e}re\surnameend} (\bibinfo{year}{1996}):
  \emph{\bibinfo{title}{Activated forms of {MMP2} and {MMP9} in abdominal
  aortic aneurysms}}.
\newblock {\sl \bibinfo{journal}{Journal of vascular surgery}}
  \bibinfo{volume}{24}(\bibinfo{number}{1}), pp. \bibinfo{pages}{127--133},
  \doi{10.1016/S0741-5214(96)70153-2}.

\bibitemdeclare{inproceedings}{92}
\bibitem{92}
\bibinfo{author}{R.~G. \surnamestart Sanfelice\surnameend},
  \bibinfo{author}{D.~A. \surnamestart Copp\surnameend} \&
  \bibinfo{author}{P.~\surnamestart Nanez\surnameend} (\bibinfo{year}{2013}):
  \emph{\bibinfo{title}{A Toolbox for Simulation of Hybrid Systems in
  {M}atlab/{S}imulink: {H}ybrid {E}quations ({HyEQ}) {T}oolbox}}.
\newblock In: {\sl \bibinfo{booktitle}{Proceedings of Hybrid Systems:
  Computation and Control Conference}}, pp. \bibinfo{pages}{101--106},
  \doi{10.1145/2461328.2461346}.

\bibitemdeclare{article}{95}
\bibitem{95}
\bibinfo{author}{L.~\surnamestart Shelton\surnameend} \& \bibinfo{author}{J.~S.
  \surnamestart Rada\surnameend} (\bibinfo{year}{2007}):
  \emph{\bibinfo{title}{Effects of cyclic mechanical stretch on extracellular
  matrix synthesis by human sclera fibroblasts}}.
\newblock {\sl \bibinfo{journal}{Experimental Eye Research}}
  \bibinfo{volume}{84}, pp. \bibinfo{pages}{314--322},
  \doi{10.1016/j.exer.2006.10.004}.

\bibitemdeclare{article}{98}
\bibitem{98}
\bibinfo{author}{S.~L. \surnamestart Shelton\surnameend} (\bibinfo{year}{July
  2009}): \emph{\bibinfo{title}{Characterization of Mechanisms Regulating
  Scleral Extracellular Matrix Remodeling to Promote Myopia Development}}.
\newblock {\sl \bibinfo{journal}{PhD thesis, University of Oklahoma Health
  Science Center, Oklahoma City}}.

\bibitemdeclare{unpublished}{Shu.Sanfelice.13.I&C}
\bibitem{Shu.Sanfelice.13.I&C}
\bibinfo{author}{Q.~\surnamestart Shu\surnameend} \& \bibinfo{author}{R.~G.
  \surnamestart Sanfelice\surnameend} (\bibinfo{year}{2013}):
  \emph{\bibinfo{title}{Dynamical Properties of a Two-gene Network with
  Hysteresis}}.
\newblock \bibinfo{note}{Submitted to Special Issue on Hybrid Systems and
  Biology, Elsevier Information and Computation}.

\bibitemdeclare{article}{siegwart2005selective}
\bibitem{siegwart2005selective}
\bibinfo{author}{J.~T. \surnamestart Siegwart\surnameend} \&
  \bibinfo{author}{T.~T. \surnamestart Norton\surnameend}
  (\bibinfo{year}{2005}): \emph{\bibinfo{title}{Selective regulation of {MMP}
  and {TIMP} {mRNA} levels in tree shrew sclera during minus lens compensation
  and recovery}}.
\newblock {\sl \bibinfo{journal}{Investigative ophthalmology \& visual
  science}} \bibinfo{volume}{46}(\bibinfo{number}{10}), pp.
  \bibinfo{pages}{3484--3492}, \doi{10.1167/iovs.05-0194}.

\bibitemdeclare{article}{97}
\bibitem{97}
\bibinfo{author}{J.~C.~H. \surnamestart Tan\surnameend}, \bibinfo{author}{F.~B.
  \surnamestart Kalapesi\surnameend} \& \bibinfo{author}{M.~T. \surnamestart
  Coroneo\surnameend} (\bibinfo{year}{2006}):
  \emph{\bibinfo{title}{Mechanosensitivity and the eye: cells coping with the
  pressure}}.
\newblock {\sl \bibinfo{journal}{British Journal of Ophthalmology}}
  \bibinfo{volume}{90}, pp. \bibinfo{pages}{383--388},
  \doi{10.1136/bjo.2005.079905}.

\bibitemdeclare{article}{xiong2006effects}
\bibitem{xiong2006effects}
\bibinfo{author}{W.~\surnamestart Xiong\surnameend},
  \bibinfo{author}{R.~\surnamestart Knispel\surnameend},
  \bibinfo{author}{J.~\surnamestart Mactaggart\surnameend} \&
  \bibinfo{author}{B.~T. \surnamestart Baxter\surnameend}
  (\bibinfo{year}{2006}): \emph{\bibinfo{title}{Effects of tissue inhibitor of
  metalloproteinase 2 deficiency on aneurysm formation}}.
\newblock {\sl \bibinfo{journal}{Journal of vascular surgery}}
  \bibinfo{volume}{44}(\bibinfo{number}{5}), pp. \bibinfo{pages}{1061--1066},
  \doi{10.1016/j.jvs.2006.06.036}.

\bibitemdeclare{article}{yang2009myopia}
\bibitem{yang2009myopia}
\bibinfo{author}{Y.~\surnamestart Yang\surnameend},
  \bibinfo{author}{X.~\surnamestart Li\surnameend},
  \bibinfo{author}{N.~\surnamestart Yan\surnameend},
  \bibinfo{author}{S.~\surnamestart Cai\surnameend} \&
  \bibinfo{author}{X.~\surnamestart Liu\surnameend} (\bibinfo{year}{2009}):
  \emph{\bibinfo{title}{Myopia: A collagen disease?}}
\newblock {\sl \bibinfo{journal}{Medical hypotheses}}
  \bibinfo{volume}{73}(\bibinfo{number}{4}), pp. \bibinfo{pages}{485--487},
  \doi{10.1016/j.mehy.2009.06.020}.

\end{thebibliography}
\end{document}